\documentclass{moriond}

\def\Journal#1#2#3#4{{#1} {\bf #2}, #3 (#4)}

\def\NIMA{{\em Nucl. Instrum. Methods} A}

\def\PRL{\em Phys. Rev. Lett.}
\def\PRD{{\em Phys. Rev.} D}

\def\be{\begin{equation}}
\def\ee{\end{equation}}
\def\bea{\begin{eqnarray}}
\def\eea{\end{eqnarray}}

\usepackage[]{units}
\usepackage{amsmath,latexsym}

\newcommand{\Photo}{}

\begin{document}
\vspace*{4cm}
\title{INCLUSIVE DETERMINATION OF $|V_{cb}|$ AT BELLE}

\author{RAYNETTE VAN TONDER \\for the Belle Collaboration}

\address{University of Bonn, Physikalisches Institut, \\Nussallee 12,
53115, Bonn, Germany}

\maketitle\abstracts{
We present the measurement of the f\mbox{}irst to fourth order moments of the four-momentum transfer squared, $q^{2}$, of inclusive $B \rightarrow X_{c} \ell \nu$ decays using the full Belle data set consisting of $\unit[711]{fb^{-1}}$ of integrated luminosity at the $\Upsilon(4S)$ resonance for $\ell = e, \mu$. The measured moments are crucial experimental inputs for a novel, data-driven approach to determine inclusive $|V_{cb}|$, which we implement in order to extract a preliminary value of $|V_{cb}| \times 10^{3} = 41.7 \pm 1.2$.}

\section{Introduction}

Precise measurements of the absolute value of the Cabibbo-Kobayashi-Maskawa (CKM) matrix element $V_{cb}$ are important to deepen our understanding of the Standard Model of Particle Physics (SM)\,\cite{Cabibo,KM}. The current world averages of $V_{cb}$ from exclusive and inclusive determinations are\,\cite{hflav}:
\bea
|V_{cb}^{\textrm{excl.}}| &=& (39.25 \pm 0.56) \times 10^{-3}~, \nonumber \\
|V_{cb}^{\textrm{incl.}}| &=& (42.19 \pm 0.78) \times 10^{-3}~,
\label{eq:sp}
\eea
where the uncertainties are the sum from experiment and theory. Both world averages exhibit a disagreement of approximately 3 standard deviations with one another.

Inclusive determinations of $|V_{cb}|$ exploit the fact that the total decay rate can be expanded into a manageable set of non-perturbative matrix elements using the Heavy Quark Expansion (HQE). A novel and alternative approach\,\cite{keri} to determine $|V_{cb}|$ from inclusive decays exploits reparametrization invariance that reduces the full set of 13 non-perturbative matrix elements present in the total rate to a set of only 8 parameters at the order of $\mathcal{O}(1/m_{b}^{4})$. The key prerequisite giving rise to the reparametrization invariance in the total rate holds only for the moments of the four-momentum transfer squared, $q^{2}$, of $B \rightarrow X_{c} \ell \nu$ decays. 
The f\mbox{}irst measurement of the f\mbox{}irst moment of the $q^{2}$ spectrum was reported by CLEO\,\cite{cleo} with a lepton energy requirement of $\unit[1]{GeV}$. Here, a f\mbox{}irst systematic study of the f\mbox{}irst to fourth moments, $\langle q^{2} \rangle$, $\langle q^{4} \rangle$, $\langle q^{6} \rangle$, $\langle q^{8} \rangle$ without a lower lepton energy cut and using a progression of cuts \mbox{$q^{2} \in [3.6,10.1]$ GeV$^{2}$} is presented.

\section{Event Reconstruction}
The full Belle data set of ($772 \pm 10$) $\times 10^{6}$ $B$ meson pairs, that were produced at the KEKB accelerator complex\,\cite{Kekb} with a centre-of-mass energy of $\sqrt{s} = \unit[10.58]{GeV}$, is analysed for this measurement. Monte Carlo (MC) samples of $B$ meson decays and continuum processes ($e^{+}e^{-} \rightarrow q\bar{q}$ with $q = u, d, s, c$) are simulated using the EvtGen generator\,\cite{evtgen}, while the detector response is modelled with GEANT3\,\cite{geant}. The modelling of inclusive semileptonic $B \rightarrow X_{c} \ell \nu$ signal decays is described in detail elsewhere\,\cite{Lu}.

Collision events are reconstructed using the Full Reconstruction algorithm\,\cite{FR}, which reconstructs one of the two $B$ mesons by making use of hadronic decay channels. In total 1104 decay cascades are reconstructed, resulting in an ef\mbox{}f\mbox{}iciency\,\cite{Bevan} of 0.28\% and 0.18\% for charged and neutral $B$ meson pairs. The output classif\mbox{}ier score of this neural network represents the estimated quality of the reconstructed candidates, denoted as the $B_{\textrm{tag}}$, and the best candidates for each event are selected. In order to suppress continuum processes the $B_{\textrm{tag}}$ candidates are required to have a beam-constrained mass of $M_{bc} = \sqrt{(s/2)^{2} - |\textbf{p}_{\textrm{tag}}|^{2}} > \unit[5.27]{GeV}$. 

After the reconstruction of the $B_{\textrm{tag}}$ candidate all remaining tracks and clusters are used to reconstruct the signal side. By using the momentum of the $B_{\textrm{tag}}$ candidate together with the precisely known initial beam-momentum, the signal $B$ rest frame is def\mbox{}ined as:
\begin{equation}
p_{\textrm{sig}} = p_{e^{+}e^{-}} - \Big(\sqrt{m_{B}^{2} + |\textbf{p}_{\textrm{tag}}|^{2}}, \textbf{p}_{\textrm{tag}}  \Big)~.
\end{equation}
Since multiple leptons are likely to originate from a double semileptonic $b \rightarrow c \rightarrow s$ cascade, exactly one signal lepton per event is required. In addition, the charge of the signal lepton is required to be opposite to that of the $B_{\textrm{tag}}$ charge. The four-momentum of the hadronic system, $p_{X}$, is reconstructed from the sum of the remaining unassigned charged particles and neutral energy depositions. Furthermore, the hadronic mass of the $X$ system is reconstructed as $M_{X} = \sqrt{(p_{X})^{\mu}(p_{X})_{\mu}}$~. With the $X$ system reconstructed, the four-momentum of the neutrino in the event is estimated by calculating the missing four-momentum,
\begin{equation}
P_{\textrm{miss}} = (E_{\textrm{miss}}, \textbf{p}_{\textrm{miss}}) = p_{\textrm{sig}} - p_{X} - p_{\ell}~.
\end{equation}
For correctly reconstructed semileptonic $B \rightarrow X_{c} \ell \nu$  decays $E_{\textrm{miss}} \approx |\textbf{p}_{\textrm{miss}}|$ and events with $E_{\textrm{miss}} - |\textbf{p}_{\textrm{miss}}| \in [-0.5,0.5]$ GeV are selected. Finally, the four-momentum transfer squared is reconstructed as $q^{2} = (p_{\textrm{sig}} - p_{X})^{2}$.

\section{Calculation of Moments}
\begin{figure}
\centering
    \includegraphics[width=0.85\linewidth]{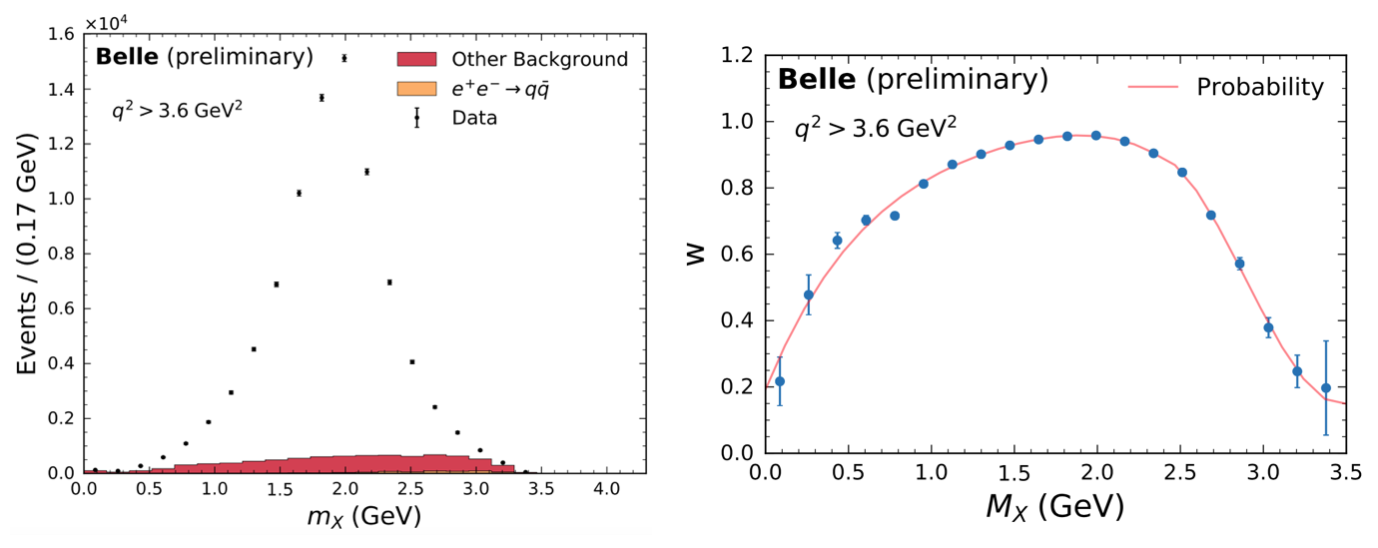}
    \caption[]{The reconstructed $M_{X}$ distribution for $q^{2} > \unit[3.6]{GeV^{2}}$ (left) and the corresponding binned signal probabilities (right).}
\label{fig:bkg}
\end{figure}
In order to subtract background events a two step procedure is carried out. First, a binned likelihood f\mbox{}it of the $M_{X}$ distribution is performed to determine the number of expected signal and background events. The signal and background templates for the f\mbox{}it are determined from MC samples, while systematic uncertainties are incorporated via nuisance-parameter constraints. Next, the determined number of background events is used to construct binned signal probabilities def\mbox{}ined as 
\begin{equation}
    w_{i} = \frac{N_{\text{Total}}^{i} - N_{\text{Bkg}}^{i}}{N_{\text{Total}}^{i}}~.
\end{equation}
The binned signal probabilities for each $q^{2}$ selection are f\mbox{}itted with a polynomial least-square f\mbox{}it of a given order $n$ to determine continuous event-wise weights, $w(M_{X})$. Figure~\ref{fig:bkg} shows the $M_{X}$ spectrum and the $w_{i}$ distribution for the selection with $q^{2} > \unit[3.6]{GeV^{2}}$.

The reconstructed $q^{2}$ values are corrected from reconstruction and selection ef\mbox{}fects by performing three sequential steps: f\mbox{}irstly, an event-wise calibration function is applied as a function of the reconstructed $q^{2}$ value to correct for resolution distortions. This linear calibration function is determined by comparing the reconstructed and generator-level moments and is derived separately for each order of the moments we wish to measure. Subsequently, the calibrated $q^{2}$ values are calculated by inverting the calibration curve for each order $m$: $q^{2m}_{\text{cal}} = (q^{2m} - a_{m})/b_{m}$, with $a_{m}$ and $b_{m}$ denoting the intercept and slope.
\begin{figure}
\centering
    \includegraphics[width=0.85\linewidth]{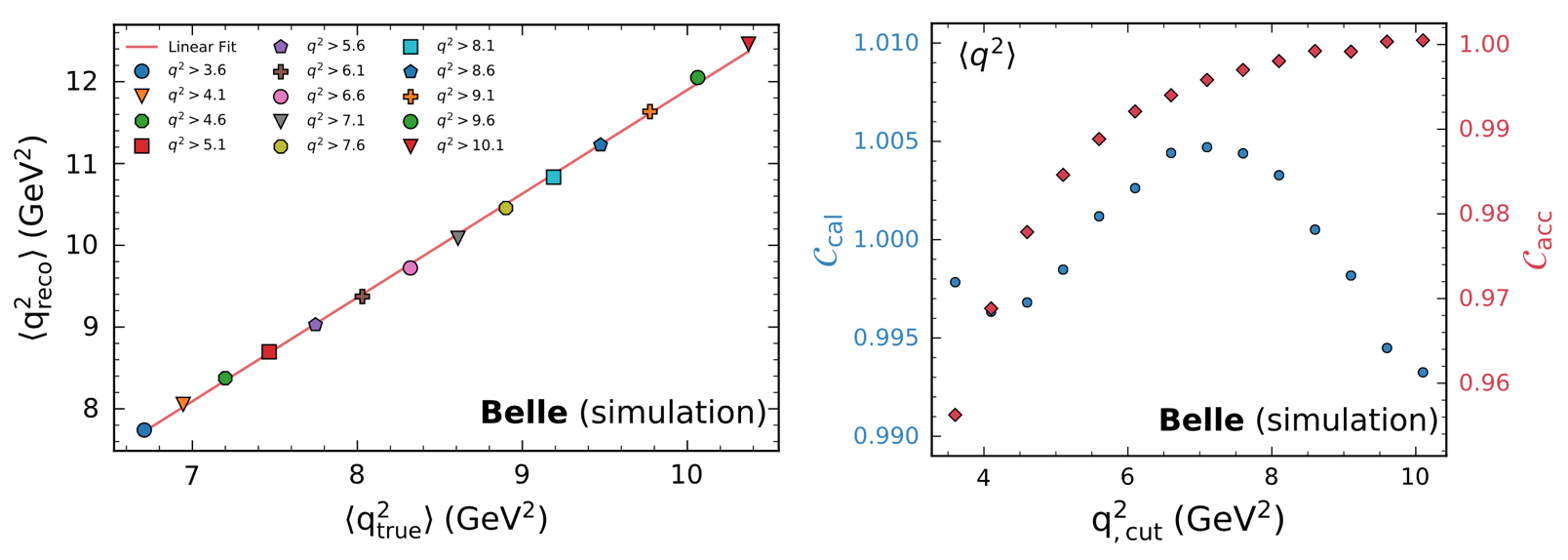}
    \caption[]{The generator-level and reconstructed $B \rightarrow X_{c} \ell \nu$ moments for the f\mbox{}irst moment for dif\mbox{}ferent $q^{2}$ selections (left), together with the residual bias correction factors $\mathcal{C}_{\text{cal}}$ and $\mathcal{C}_{\text{acc}}$ (right). }
\label{fig:calib}
\end{figure}
Subsequently, residual bias that may arise from small non-linearities in the extracted calibration curves is corrected with a global correction factor, denoted as $\mathcal{C}_{\text{cal}}$. Lastly, acceptance and selection losses are corrected by applying a global correction factor, $\mathcal{C}_{\text{acc}}$. The residual bias correction factor $\mathcal{C}_{\text{cal}}$ is determined by comparing the calibrated and generator-level moments for each order and $q^{2}$ selection, while $\mathcal{C}_{\text{acc}}$ is determined by comparing the generator-level moments with a sample without any selection criteria. The linear calibration functions as well as the additional correction factors are determined from MC samples that are statistically independent from the samples used in the background subtraction procedure and are shown in Figure~\ref{fig:calib} for the f\mbox{}irst moment. Finally, the $q^{2}$ moments are given by
 \begin{equation}
 \langle q^{2m}\rangle = \frac{\mathcal{C}_{\text{cal}} \cdot \mathcal{C}_{\text{acc}}}{\sum^{\text{events}}_{i} w(M_{Xi})} \sum^{\text{events}}_{i} w(M_{Xi}) \cdot q^{2m}_{\text{cal} \mbox{ }i}~.
 \end{equation}
\section{Results}
\begin{figure}
\centering
    \includegraphics[width=0.77\linewidth]{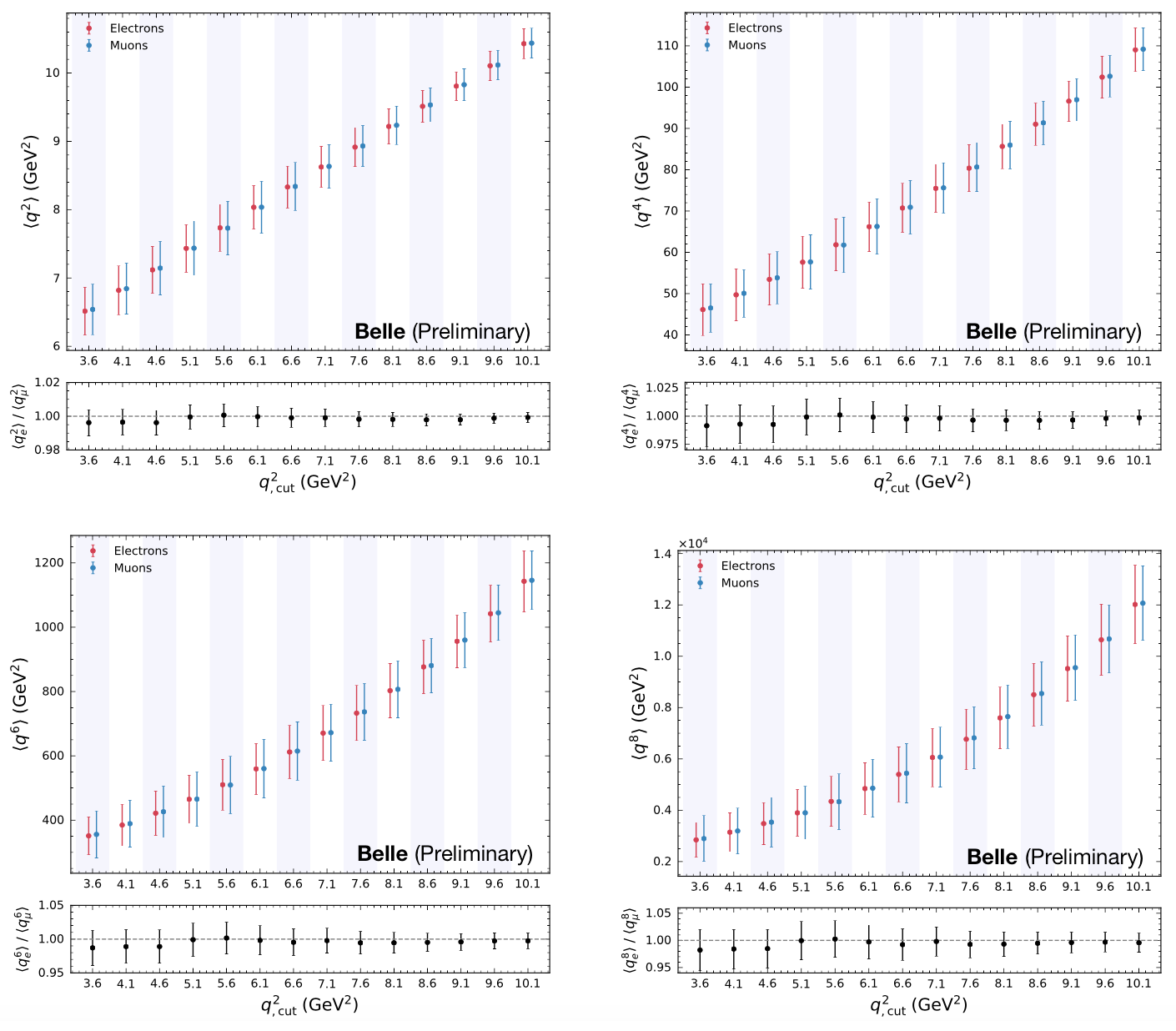} 
    \caption[]{The f\mbox{}irst to fourth order measured moments for electrons and muons with the total statistical and systematic errors scaled by a factor of 10. Note that the individual electron and muon moments are strongly correlated.}
\label{fig:rmoments}
\end{figure}

Figure~\ref{fig:rmoments} shows the f\mbox{}irst to fourth order measured $q^{2}$ moments for electrons and muons. In the shown ratio many of the associated systematic uncertainties cancel and all reported moments are compatible with the expectation of lepton f\mbox{}lavour universality.  

A $\chi^{2}$-f\mbox{}it is performed\,\cite{kevin} to the combined measured moments using a conservative estimate for theoretical uncertainties in order to extract a value of $|V_{cb}| \times 10^{3} = 41.7 \pm 1.2$.

\section{Outlook}
Preliminary measurements of the f\mbox{}irst to the fourth moments of the $B \rightarrow X_{c} \ell \nu$ $q^{2}$ spectrum with several selections on $q^{2}$ are reported. The measured moments are crucial experimental inputs for a novel, data-driven approach to determine inclusive $|V_{cb}|$, which we implement in order to extract a f\mbox{}irst preliminary value. The f\mbox{}inal analysis will incorporate a few modif\mbox{}ications including additional systematics regarding signal and background modelling, as well as improving the calculation of the background subtraction functions. Additionally, a dif\mbox{}ferent set of $q^{2}$ selections will be considered for the f\mbox{}inal measured moments.

\section*{Acknowledgments}
This research was supported by the DFG Emmy-Noether Grant No. BE 6075/1-1.

\end{document}